\documentclass[aps,prd,twocolumn,superscriptaddress,floatfix]{revtex4}
\usepackage{bm}
\usepackage{graphicx}
\usepackage{amsmath,amssymb,times}

\newcommand{\bequ}{\begin{equation}}
\newcommand{\eequ}{\end{equation}}
\newcommand{\bea}{\begin{eqnarray}}
\newcommand{\eea}{\end{eqnarray}}

\DeclareSymbolFont{boldletters}{OML}{cmm} {b}{it}
\DeclareSymbolFontAlphabet{\mathbit}{boldletters}
\DeclareMathSymbol{\alpha}{\mathalpha}{letters}{"0B}
\DeclareMathSymbol{\beta}{\mathalpha}{letters}{"0C}
\DeclareMathSymbol{\gamma}{\mathalpha}{letters}{"0D}
\DeclareMathSymbol{\delta}{\mathalpha}{letters}{"0E}
\DeclareMathSymbol{\epsilon}{\mathalpha}{letters}{"0F}
\DeclareMathSymbol{\zeta}{\mathalpha}{letters}{"10}
\DeclareMathSymbol{\eta}{\mathalpha}{letters}{"11}
\DeclareMathSymbol{\theta}{\mathalpha}{letters}{"12}
\DeclareMathSymbol{\iota}{\mathalpha}{letters}{"13}
\DeclareMathSymbol{\kappa}{\mathalpha}{letters}{"14}
\DeclareMathSymbol{\lambda}{\mathalpha}{letters}{"15}
\DeclareMathSymbol{\mu}{\mathalpha}{letters}{"16}
\DeclareMathSymbol{\nu}{\mathalpha}{letters}{"17}
\DeclareMathSymbol{\xi}{\mathalpha}{letters}{"18}
\DeclareMathSymbol{\pi}{\mathalpha}{letters}{"19}
\DeclareMathSymbol{\rho}{\mathalpha}{letters}{"1A}
\DeclareMathSymbol{\sigma}{\mathalpha}{letters}{"1B}
\DeclareMathSymbol{\tau}{\mathalpha}{letters}{"1C}
\DeclareMathSymbol{\upsilon}{\mathalpha}{letters}{"1D}
\DeclareMathSymbol{\phi}{\mathalpha}{letters}{"1E}
\DeclareMathSymbol{\chi}{\mathalpha}{letters}{"1F}
\DeclareMathSymbol{\psi}{\mathalpha}{letters}{"20}
\DeclareMathSymbol{\omega}{\mathalpha}{letters}{"21}
\DeclareMathSymbol{\varepsilon}{\mathalpha}{letters}{"22}
\DeclareMathSymbol{\vartheta}{\mathalpha}{letters}{"23}
\DeclareMathSymbol{\varpi}{\mathalpha}{letters}{"24}
\DeclareMathSymbol{\varrho}{\mathalpha}{letters}{"25}
\DeclareMathSymbol{\varsigma}{\mathalpha}{letters}{"26}
\DeclareMathSymbol{\varphi}{\mathalpha}{letters}{"27}
\DeclareMathSymbol{\Gamma}{\mathalpha}{letters}{"00}
\DeclareMathSymbol{\Delta}{\mathalpha}{letters}{"01}
\DeclareMathSymbol{\Theta}{\mathalpha}{letters}{"02}
\DeclareMathSymbol{\Lambda}{\mathalpha}{letters}{"03}
\DeclareMathSymbol{\Xi}{\mathalpha}{letters}{"04}
\DeclareMathSymbol{\Pi}{\mathalpha}{letters}{"05}
\DeclareMathSymbol{\Sigma}{\mathalpha}{letters}{"06}
\DeclareMathSymbol{\Upsilon}{\mathalpha}{letters}{"07}
\DeclareMathSymbol{\Phi}{\mathalpha}{letters}{"08}
\DeclareMathSymbol{\Psi}{\mathalpha}{letters}{"09}
\DeclareMathSymbol{\Omega}{\mathalpha}{letters}{"0A}

\def\fun#1#2{\lower3.6pt\vbox{\baselineskip0pt\lineskip.9pt
\ialign{$\mathsurround=0pt#1\hfil##\hfil$\crcr#2\crcr\sim\crcr}}}

\begin{document}
\title{
Pion and $\rho$-meson screening masses at finite chemical potential\\
in two-flavor lattice QCD with Wilson fermion
}

\author{Junpei Sugano}
\email[]{sugano@phys.kyushu-u.ac.jp}
\affiliation{Department of Physics, Graduate School of Sciences, Kyushu University,
             Fukuoka 819-0395, Japan}

\author{Junichi Takahashi}
%\email[]{j.t.mhjkk.f.c@gmail.com}
\affiliation{Division of Observation, Fukuoka Aviation Weather Station,
Japan Meteorological Agency, Fukuoka 812-0005, Japan}

\author{Hiroaki Kouno}
%\email[]{kounoh@cc.saga-u.ac.jp}
\affiliation{Department of Physics, Saga University,
             Saga 840-8502, Japan}

\author{Masanobu Yahiro}
%\email[]{yahiro@phys.kyushu-u.ac.jp}
\affiliation{Department of Physics, Graduate School of Sciences, Kyushu University,
             Fukuoka 819-0395, Japan}

\date{\today}

%%%%%%%%%%%%%%%%%%%%%%%%%%%%%%%%%%%%%%%%%%%%%%%%%%%%%%%%%%%%%%%%%%%%%%%%%%%
%%%%%  Abstract 
%%%%%%%%%%%%%%%%%%%%%%%%%%%%%%%%%%%%%%%%%%%%%%%%%%%%%%%%%%%%%%%%%%%%%%%%%%%
\begin{abstract}
 We investigate the real and the imaginary chemical-potential ($\mu$) 
 dependence of pion and $\rho$-meson screening masses
 in both the confinement and the deconfinement region
 by using two-flavor lattice QCD.
 The spatial meson correlators are calculated in the imaginary $\mu$ region 
 with lattice QCD simulations on an $8^{2}\times 16\times 4$ lattice
 with the clover-improved two-flavor Wilson fermion action
 and the renormalization-group-improved Iwasaki gauge action.
 We extract pion and $\rho$-meson screening masses
 from the correlators. 
 The meson screening masses thus obtained are extrapolated
 to the real $\mu$ region by assuming either the Fourier or 
 the polynomial series. 
 In the real $\mu$ region, 
 the resulting pion and $\rho$-meson screening masses monotonically increase 
 as real $\mu$ becomes large.
\end{abstract}

%pacs{11.15.Ha, 12.38.Gc, 14.40.-n}
\maketitle
%%%%%%%%%%%%%%%%%%%%%%%%%%%%%%%%%%%%%%%%%%%%%%%%%%%%%%%%%%%%%%%%%%%%%%%%%%%
%%%%%  Introduction 
%%%%%%%%%%%%%%%%%%%%%%%%%%%%%%%%%%%%%%%%%%%%%%%%%%%%%%%%%%%%%%%%%%%%%%%%%%%
\section{Introduction}
Understanding of the QCD phase diagram
\cite{Munzinger-Wambach, Fukushima-Hatsuda, Fukushima-Sasaki}
is a long-standing issue in hadron physics.
The knowledge of thermal properties of the QCD
is essential to clarify the phase diagram,
and lattice QCD (LQCD) simulations are well established
as a powerful tool.
Indeed, LQCD simulations are successful in
clarifying the phase diagram and the properties of QCD
at zero chemical potential $(\mu)$ and finite temperature $(T)$
~\cite{Borsanyi}.
It is, however, difficult to perform LQCD simulations
for finite real $\mu$, since
the fermion determinant $\textrm{det}\mathcal{M}(\mu)$ becomes complex:
\begin{align}
 (\textrm{det}\mathcal{M}(\mu))^{*}=\textrm{det}\mathcal{M}(-\mu^{*})
  =\textrm{det}\mathcal{M}(-\mu).
  \label{sign-problem}
\end{align}
This is the well-known sign problem.
It prevents us from using Monte-Carlo methods
based on the importance sampling. 

Several methods were proposed so far, in order to
circumvent the sign problem~\cite{Forcrand_sign_problem},
e.g.,
the Taylor expansion method~\cite{Allton1, Allton2, Ejiri},
the reweighting method~\cite{Fodor-Katz-reweighting}, 
the analytic continuation from the purely-imaginary $\mu$ region
to the real $\mu$ region~\cite{Forcrand-Philipsen, DElia-Lombardo},
and
the canonical approach~\cite{Muroya,Nakamura,Bornyakov}.
Recently,
the complex Langevin method~\cite{Aarts1, Aarts2, Sexty, Aarts3}
and the Lefschetz thimble theory~\cite{Cristoforetti, Fujii}
were proposed as the new methods, and made a great progress.
Among these methods,
we focus on the imaginary $\mu$ approach in this paper.
For purely-imaginary chemical potential $\mu=i\mu_{\rm I}=i\theta T$,
the first equality of Eq.~(\ref{sign-problem})
ensures that the fermion determinant $\textrm{det}\mathcal{M}(i\theta T)$ 
is real. 
Here, $\theta$ is a dimensionless chemical potential.
This means that LQCD simulations can be performed
with usual Monte-Carlo methods
for finite $\theta$.
Observables calculated at $\theta=\mu_{\rm I}/T$
are analytically continued
to real $\mu/T$ ($\mu_{\rm R}/T$), by assuming that
the $\theta$ dependence of observables can be described
by some analytic function.

Toward the clarification of the QCD phase diagram,
meson screening masses are extensively calculated by using
LQCD simulations~\cite{Umeda,Cheng:2010fe,Bazavov}.
Indeed, the meson screening masses are good
indicators to see chiral and $U_{\rm A}(1)$ symmetry restorations
~\cite{Ishii1, Ishii2},
and hence essential quantities to explore the QCD phase diagram.
It is also expected that the meson screening masses play a key role
in investigating medium properties of hadronic excitations 
in the Quark Gluon Plasma~\cite{Adams} that may be created 
by relativistic heavy-ion collision experiments.
As for finite $\mu_{\rm R}$,
pion and $\rho$-meson screening masses were calculated
up to order $(\mu_{\rm R}/T)^2$
with the Taylor expansion method for 
both staggered-type fermions~\cite{QCD-TARO} 
and Wilson-type ones~\cite{Iida}.

In this paper, we investigate the $\mu$ dependence of 
pion and $\rho$-meson screening masses
in both the imaginary and the real $\mu$ region 
by using two-flavor LQCD simulations.
We first calculate the spatial
pion and $\rho$-meson correlators
in the purely-imaginary $\mu$ region, i.e., the $\theta$ region.
The simulations can be made
with standard numerical prescriptions,
since there is no sign problem in the $\theta$ region.
The calculated correlators are fitted by
the exponential form at large distance,
in order to derive the screening masses.
To perform the analytic continuation
from the imaginary to the real $\mu$ region,
we fit the resulting meson screening masses
by the Fourier or the polynomial series in the $\theta$ region.
After the fitting,
the meson screening masses at finite $\mu_{\rm R}/T$
are extracted by taking 
the replacement $\theta \rightarrow -i\mu_{\rm R}/T$
in the series.

Actual LQCD simulations are done on an $8^{2}\times 16 \times 4$ lattice
with the clover-improved two-flavor Wilson fermion action
and the renormalization-group-improved Iwasaki gauge action.
We adopt the line of constant physics with 
$m_{\mathrm{PS}}/m_{\mathrm{V}}=0.80$
obtained in Refs.~\cite{Maezawa,Khan1,Khan2}
for finite-temperature simulations,
where $m_{\mathrm{PS}}$ and $m_{\mathrm{V}}$ are 
pseudoscalar-meson and vector-meson masses, respectively.
Three temperatures $T/T_{\rm pc}=0.93, 1.08, $ and 1.35 
are considered.
The pseudocritical
temperature at $\mu=0$ is represented by $T_{\rm pc}$~\cite{Ejiri,Khan1}.
We compute spatial pion and $\rho$-meson correlators
at these temperatures and in the range
$0\le \theta \le \pi/3$.
We generated about 32,000 trajectories
and removed the first 4,000 trajectories
for the thermalization of $T$ and $\theta$,
and then measured pion and $\rho$-meson correlators
at every 100 trajectories. 
Lattice gauge configurations taken above are the same as in 
our previous work~\cite{takahashi2} 
where the quark number density was analyzed.

The rest of this paper is organized as follows. 
In Sec.~\ref{Formulation}, we explain
the meson screening mass and the analytic continuation.
In Sec.~\ref{Numerical results}, we show
numerical results for the meson screening masses at both imaginary and real $\mu$.  
Section~\ref{Summary} is devoted to a summary.

%%%  Formulation
\section{Formulation}
\label{Formulation}

In this section,
we explain the formulation of
meson screening mass and
analytic continuation from the
$\theta$ region to the $\mu_{\rm R}/T$ region.
As for LQCD setup, see Ref.~\cite{takahashi2}.

\subsection{Meson screening mass}
We extract pion and $\rho$-meson screening masses
at finite $\theta$
from the spatial correlator
\begin{eqnarray}
 C_{i}(z;T,\theta) = \sum_{x,y,t}\langle M_{i}(x,y,z,t)
  M_{i}^{\dag}(0,0,0,0) \rangle,
  \label{correlator}
\end{eqnarray}
with the meson operator
\begin{eqnarray}
 M_{i}(x,y,z,t)\equiv \bar{q}(x,y,z,t)\Gamma_{i} \tau^{a} q(x,y,z,t),
\end{eqnarray}
where the subscript 
$i$ represents the species of meson,
$\tau^{a}$ is the Pauli matrix
in flavor space, and
$\Gamma_{\pi}=\gamma_{5}$ for pion
and $\Gamma_{\rho}=\gamma_{\mu}$ 
for $\rho$-meson.
The correlator is summed over $x,y,t$
in order to project
on zero momenta in the $x$- and $y$-directions
and on zero energy in the $t$-direction.

Considering large $z$, we derive the meson screening mass $m_{i}(T,\theta)$
from $C_{i}(z;T,\theta)$ by fitting it with the exponential form: 
\begin{eqnarray}
 C_{i}(z;T,\theta) = A_{i}(T,\theta)
  \left(
   \textrm{e}^{-m_{i}(T,\theta)z}
   +\textrm{e}^{-m_{i}(T,\theta)(N_{z}-z)}
  \right),
\end{eqnarray}
where $A_{i}(T,\theta)$ is the fitting parameter
together with $m_{i}(T,\theta)$,
and $N_{z}$ is the lattice size in the $z$-direction.
The correlator (\ref{correlator}) is charge-even.
This ensures that $m_{i}(T,\theta)$ is
also charge-even and real,
even when the chemical potential is purely imaginary.
In fact, we have confirmed
that the calculated correlator has no imaginary part.

\subsection{Analytic continuation}
\begin{figure}[t]
\begin{center}
 \includegraphics[width=0.45\textwidth]{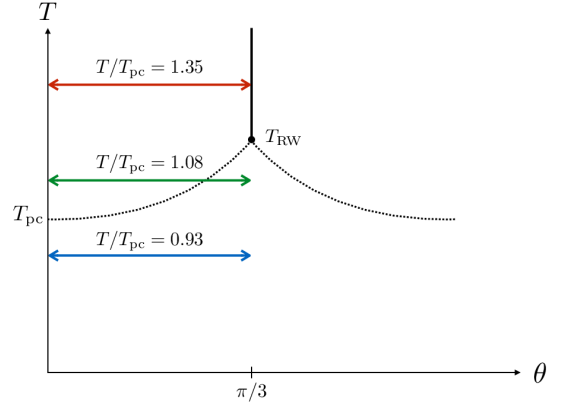}
\end{center}
 \vspace{-10pt}
 \caption{
 Sketch of the phase diagram in the $\theta$ region.
 The dotted line denotes the pseudocritical
 line, and the solid line does the
 first-order Roberge-Weiss transition line.
 The symbols $T_{\rm pc}$ and $T_{\rm RW}$
 mean the pseudocritical temperature at $\theta=0$
 and the endpoint temperature of the Roberge-Weiss transition
 at $\theta=\pi/3$, respectively.
 The arrows indicate 
 three cases of $T/T_{\rm pc}=0.93, 1.08$ and $1.35$. 
}
 \label{phase_diagram_imaginary}
\end{figure}

Our goal is to obtain the meson screening mass
at $\mu_{\rm R}/T$.
We then extrapolate the calculated $m_{i}(T,\theta)$
to the $\mu_{\rm R}/T$ region,
assuming some analytic function for each temperature taken.
In this paper, we consider three temperatures, e.g.,
$T/T_{\rm pc}=0.93, 1.08$ and $1.35$. 
Figure~\ref{phase_diagram_imaginary} shows 
the phase diagram in $T$--$\theta$ plane. 
The arrows present three cases 
of $T/T_{\rm pc}=0.93, 1.08$ and $1.35$. 
The system is in the confinement region at $T/T_{\rm pc}=0.93$,
and in the deconfinement region at $T/T_{\rm pc}=1.35$
for any $\theta$ in $0\le \theta \le \pi/3$.
As for $T/T_{\rm pc}=1.08$,
it was found in our previous work~\cite{takahashi2} that
the temperature satisfies $T_{\rm pc}<T<T_{\rm RW}$,
where $T_{\rm RW}$ corresponds to 
the endpoint of the first-order Roberge-Weiss (RW)
transition~\cite{Roberge-Weiss}. 
This means that the system changes from the deconfinement
region to the confinement one at some
value $\theta=\theta_{\rm c}$.

%%    Fig. 1

In determining the reasonable analytic function
for each temperature, the 
behavior of physical quantity $\mathcal{O}(T,\theta)$
in the $\theta$ region is essential.
For $T<T_{\rm pc}$,
it is found that 
$\mathcal{O}(T,\theta)$ is a smooth function of $\theta$
and has a periodicity of $2\pi/3$ in $\theta$
~\cite{Roberge-Weiss, Kouno, Sakai1, Sakai2, Sakai_JPhys}.
From this point of view,
we use the Fourier series for $T/T_{\rm pc}=0.93$ as
an extrapolation function,
\begin{align}
 \frac{m_{i}(T,\theta)}{T}
 =G^{n}_{\textrm{F},i}(T,\theta)
 =\sum_{k=0}^{n}a^{(k)}_{\textrm{F},i}(T)\cos(3k\theta),
 \label{theta-Fourier}
\end{align}
where $n$ denotes the highest order of series.
The function $\sin(3k\theta)$ does not appear in Eq.~(\ref{theta-Fourier})
because both $m_{\pi}$ and $m_{\rho}$ are charge-even.
Then, the screening mass in the $\mu_{\rm R}/T$ region
is obtained by the analytic continuation, that is, by the
replacement $\theta\rightarrow -i\mu_{\rm R}/T$:
\begin{align}
 \frac{m_{i}(T,\mu_{\rm R}/T)}{T}
 & = H^{n}_{\textrm{F},i}\left(T,\frac{\mu_{\rm R}}{T}\right)
 \notag \\
 & =\sum_{k=0}^{n}a^{(k)}_{\textrm{F},i}(T)\cosh\left(3k\frac{\mu_{\rm R}}{T}\right).
  \label{mu_R-Fourier}
\end{align}
Note that the coefficients $a^{(k)}_{\textrm{F},i}(T)$
in Eq.~(\ref{mu_R-Fourier})
have already been determined in the $\theta$ region.
This is true for other temperatures.

As for $T_{\rm pc}<T$,
two cases can be considered;
one is $T_{\rm RW}<T$ and the other is $T_{\rm pc}<T<T_{\rm RW}$.
For $T_{\rm RW}<T$, the 
first-order RW phase transition takes place
at $\theta=\pi/3$~\cite{Roberge-Weiss},
and analyticity of $\mathcal{O}(T,\theta)$ is lost.
Indeed, on the RW phase transition line,
a cusp comes out for charge-even $\mathcal{O}(T,\theta)$
\cite{Kouno, Sakai1, Sakai2,Sakai_JPhys}
such as $m_{i}(T,\theta)$. 
This indicates that $\mathcal{O}(T,\theta)$
monotonically increases or decreases in the region $[0,\pi/3]$.
As for $T/T_{\rm pc}=1.35$, therefore,
the polynomial series including only even powers
is applied
to extrapolate $m_{i}(T,\theta)/T$
to the $\mu_{\rm R}/T$ region:
\begin{eqnarray}
 \frac{m_{i}(T,\theta)}{T}
  =G^{n}_{\textrm{P},i}(T,\theta)
  =\sum_{k=0}^{n}a^{(k)}_{\textrm{P},i}(T)\theta^{2k}.
  \label{theta-Polynomial}
\end{eqnarray}
After the replacement $\theta\rightarrow -i\mu_{\rm R}/T$,
we can obtain
\begin{align}
 \frac{m_{i}(T,\mu_{\rm R}/T)}{T}
 & =H^{n}_{\textrm{P},i}\left(T,\frac{\mu_{\rm R}}{T}\right)
 \notag \\
 & =\sum_{k=0}^{n}(-1)^{k}a^{(k)}_{\textrm{P},i}(T)
 \left(\frac{\mu_{\rm R}}{T}\right)^{2k}.
  \label{mu_R-Polynomial}
\end{align}

For $T/T_{\rm pc}=1.08$,
the system is in the deconfinement
region $\theta\le \theta_{\rm c}$,
while the confinement is realized for $\theta>\theta_{\rm c}$.
It is thus unclear which analytic function is suitable.
Hence, for $T/T_{\rm pc}=1.08$,
we consider the region $0\le \theta\le \theta_{\rm c}$ only,
and take the polynomial series (\ref{theta-Polynomial}).
The actual value of $\theta_{\rm c}$ is determined later.

%%    Fig. 2
\begin{figure}[t]
\begin{center}
 \includegraphics[width=0.4\textwidth]{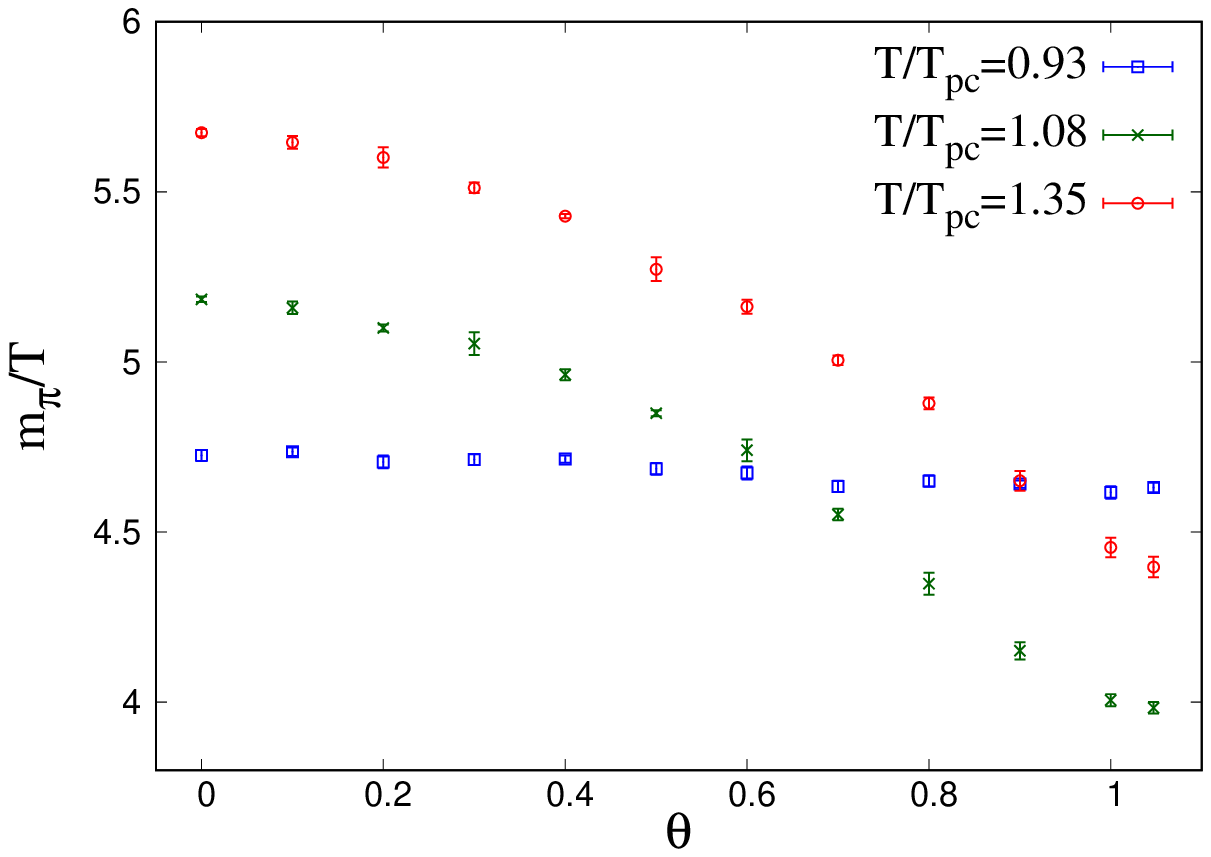}
 \includegraphics[width=0.4\textwidth]{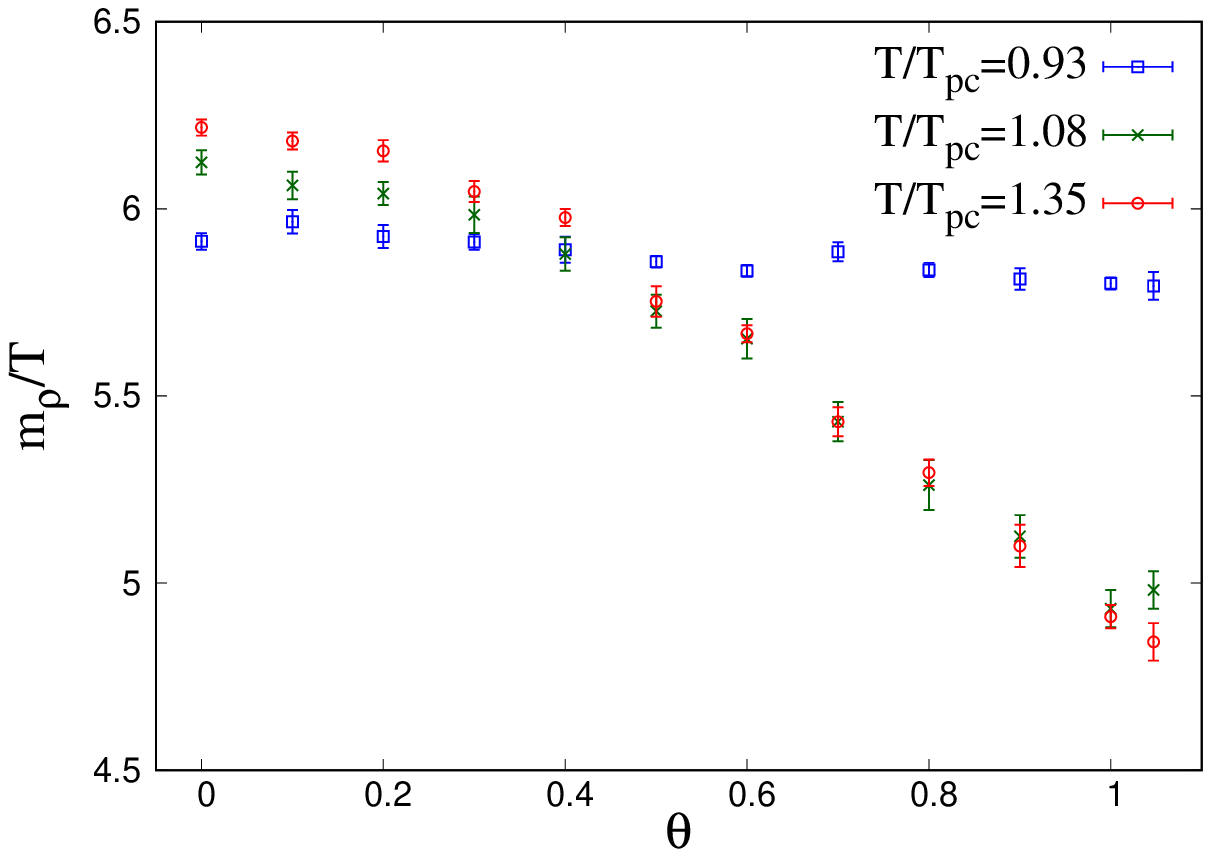}
\end{center}
\vspace{-10pt}
\caption{
The $\theta$ dependence of pion and $\rho$-meson screening masses for all 
the $T$ we consider.
The LQCD data are shown by symbols with error bars.
}
\label{meson_screening_mass_mui}
\end{figure}

%%    Fig. 3
\begin{figure}[t]
 \begin{center}
  \includegraphics[width=0.4\textwidth]{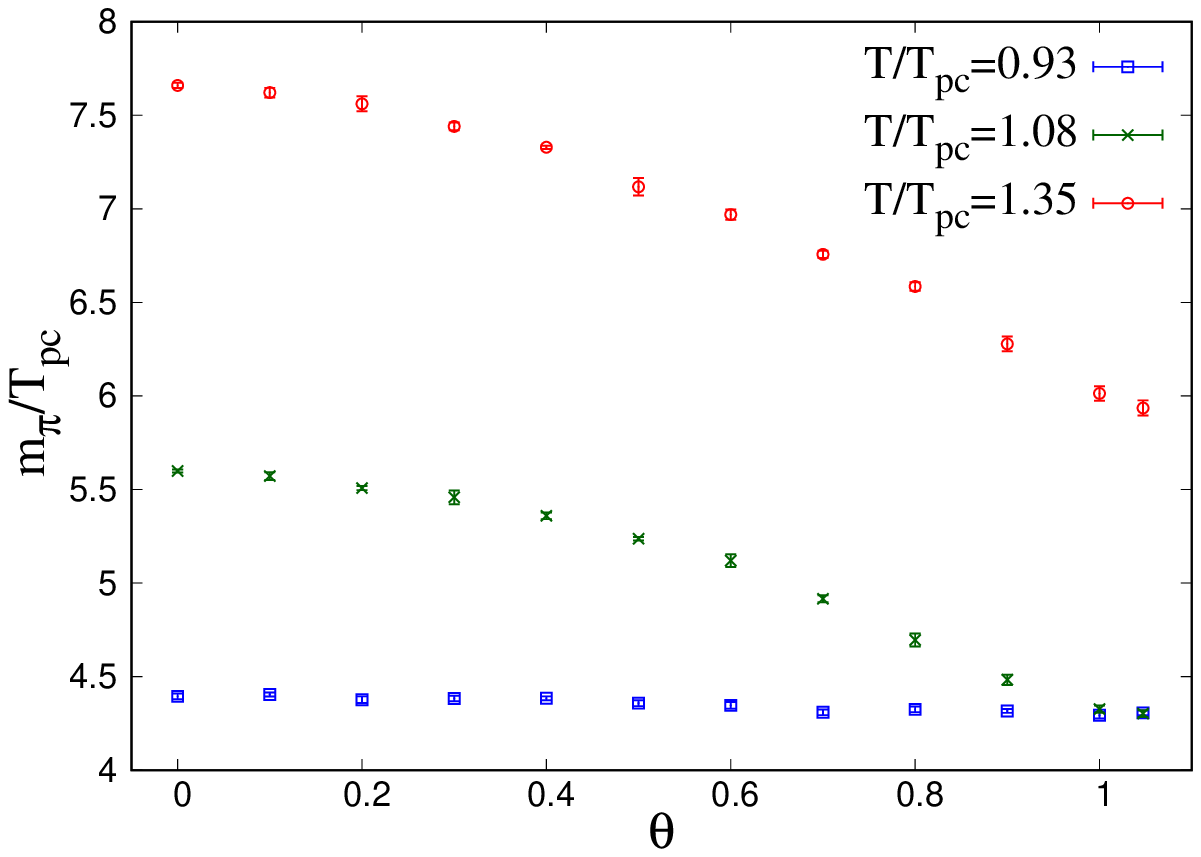}
  \includegraphics[width=0.4\textwidth]{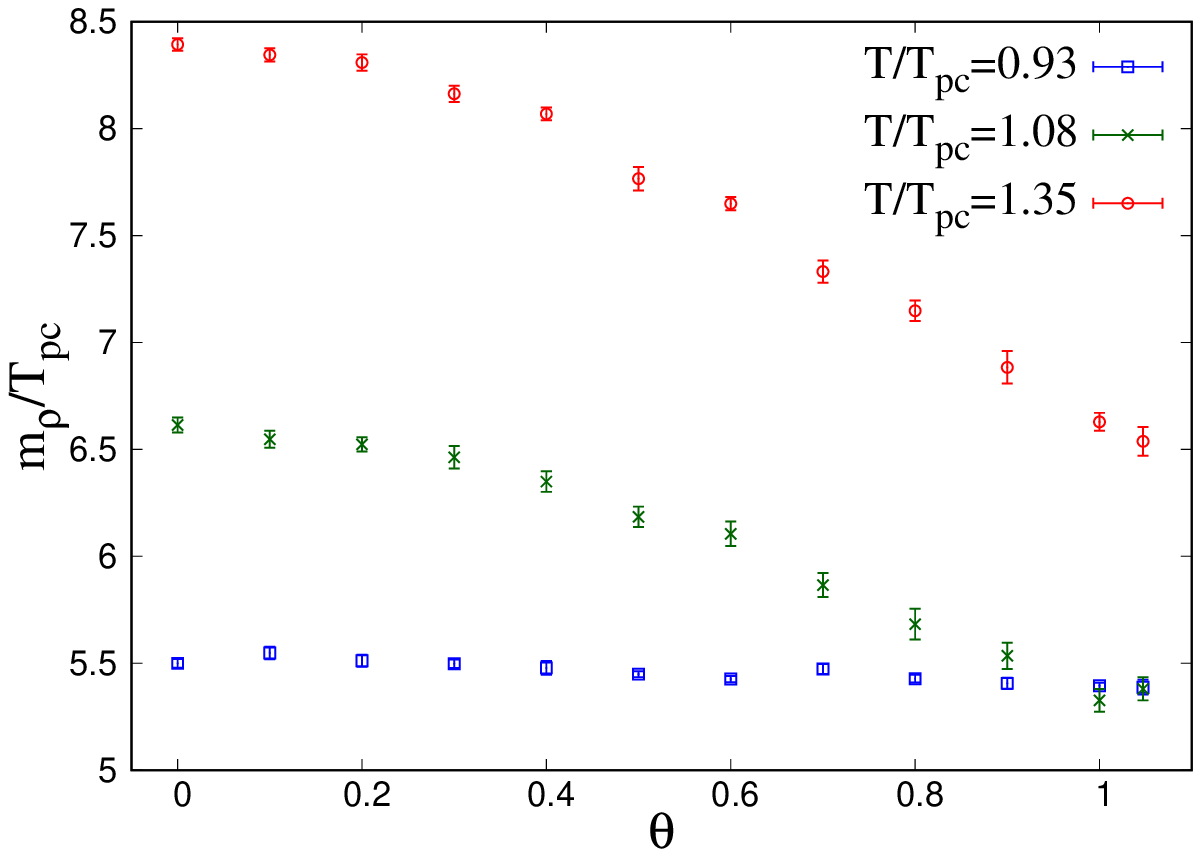}
 \end{center}
 \vspace{-10pt}
 \caption{
 The $\theta$ dependence of pion and $\rho$-meson screening masses
 devided by $T_{\rm pc}$.
 The meaning of symbols is the same as in Fig.~\ref{meson_screening_mass_mui}.
 }
 \label{meson_screening_mass_mui_GeV}
\end{figure}

%%%%%%%%  Numerical results 
\section{Numerical results}
\label{Numerical results}

\subsection{Meson screening mass at imaginary $\mu$}
\label{Sec:Meson screening mass at imaginary mu}

Figure~\ref{meson_screening_mass_mui}
shows pion and $\rho$-meson screening masses 
as a function of $\theta$ for all the $T$ we consider;
our LQCD data are plotted by symbols with error bars.
The $\theta$ dependence of the
screening masses at $T/T_{\rm pc}=0.93$ 
is quite small.
This suggests that the system is in the
chiral symmetry broken phase at $T/T_{\rm pc}=0.93$.
On the contrary, for $T/T_{\rm pc}=1.08$ and $T/T_{\rm pc}=1.35$,
the $\theta$ dependence of the screening masses
are remarkable,
and this behavior indicates realization of
the chiral symmetry restoration.

%%%%%%%%%%%%%%%%%
%%%%% TABLE 1
%%%%%%%%%%%%%%%%%
\begin{table*}[t]
 \begin{center}
  \caption{
  Coefficients of the Fourier series at $T/T_{\rm pc}=0.93$.
  }
  {\tabcolsep = 5mm
  \begin{tabular}{c|ccccc}
   \hline \hline
   meson & $T/T_{\rm pc}$ & $a^{(0)}_{\textrm{F},i}$
           & $a^{(1)}_{\textrm{F},i}$ & $a^{(2)}_{\textrm{F},i}$ & $\chi^2/{\rm dof}$ \\
   \hline
   pion & 0.93 & 4.682(4) & 0.05210(557) & ----- & 0.582 \\
        & 0.93 & 4.683(4) & 0.05170(562) & - 0.002999(5851) & 0.617 \\
   $\rho$-meson & 0.93 & 5.867(6) & 0.06103(915) & ----- & 0.897 \\
                & 0.93 & 5.867(6) & 0.06163(925) & 0.003645(7942) & 0.974 \\
   \hline \hline
  \end{tabular}
  }
  \label{TABLE_Fourier-fit}
 \end{center}
\end{table*}
%%%%%%%%%%%%%%%

%%%%%%%%%%%%%%%%%
%%%%% TABLE 2
%%%%%%%%%%%%%%%%%
\begin{table*}[t]
 \begin{center}
  \caption{
  Coefficients of the polynomial series at $T/T_{\rm pc}=1.08$.
  Note that the fitting is performed only in the range $0\le \theta \le 0.8$.
  }
  {\tabcolsep = 5mm
  \begin{tabular}{c|ccccc}
   \hline \hline
   meson & $T/T_{\rm pc}$ & $a^{(0)}_{\textrm{P},i}$
           & $a^{(1)}_{\textrm{P},i}$ & $a^{(2)}_{\textrm{P},i}$ & $\chi^2/{\rm dof}$ \\
   \hline
   pion & 1.08 & 5.171(6) & $-1.279$(26) & ----- & 1.099 \\
        & 1.08 & 5.173(6) & $-1.326$(73) & 0.1014(1474) & 1.203 \\
   $\rho$-meson & 1.08 & 6.098(19) & $-1.335$(77) & ----- & 0.340 \\
                & 1.08 & 6.103(21) & $-1.450$(247) & 0.2153(4416) & 0.357 \\
   \hline \hline
  \end{tabular}
  }
  \label{TABLE_polynomial-fit1}
 \end{center}
\end{table*}
%%%%%%%%%%%%%%%

In Fig.~\ref{meson_screening_mass_mui_GeV},
pion and $\rho$-meson screening masses
devided by $T_{\rm pc}$ are plotted as a function of $\theta$.
In both the panels, the value of screening mass at $T/T_{\rm pc}=1.08$
almost agree with that at $T/T_{\rm pc}=0.93$,
when $\theta=\theta_{\rm c}>0.8$.
This means that the system is in the deconfinement region
for $\theta\le \theta_{\rm c}$,
whereas the confinement takes place for $\theta>\theta_{\rm c}$.
Therefore, for $T/T_{\rm pc}=1.08$,
we use the data only in the range $0\le \theta\le 0.8$
for the extrapolation to the $\mu_{\rm R}/T$ region,
and the polynomial series~(\ref{theta-Polynomial})
as an extrapolation function.

%%%%%%%%%%%%%%%%%%%%%%%%%%%%%%%%%%%%%%%%%%%%%%%%%%
%%%%%%%%    Fitting of meson screening mass at imaginary $\mu$
%%%%%%%%%%%%%%%%%%%%%%%%%%%%%%%%%%%%%%%%%%%%%%%%%%
\subsection{Fitting of meson screening mass at imaginary $\mu$}
\label{Fitting of meson screening mass in imaginary mu region}

Now, we perform the $\chi^2$ fitting
for LQCD data on screening masses in the $\theta$ region.
We first consider the case of $T/T_{\rm pc}=0.93$,
and fit pion and $\rho$-meson screening masses
by the Fourier series~(\ref{theta-Fourier}).
We perform the $\chi^2$ fitting
by using $G^{1}_{\textrm{F},i}$ and $G^{2}_{\textrm{F},i}$
because
the resulting screening masses have 
small $\theta$ dependence,
as shown in Fig.~\ref{meson_screening_mass_mui}.

%%%%%%%%%%%%%%%%
%%    Fig. 4
%%%%%%%%%%%%%%%%
\begin{figure}[b]
 \begin{center}
  \includegraphics[width=0.4\textwidth]{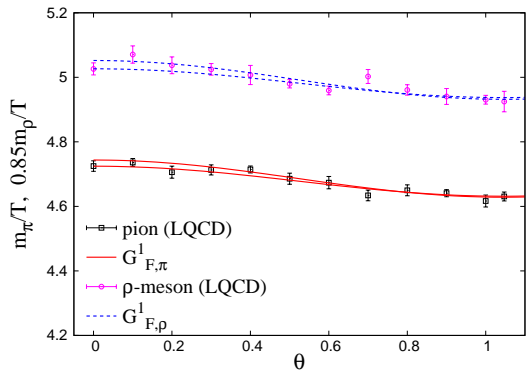}
 \end{center}
 \vspace{-10pt}
 \caption{
 Fitting results of pion and $\rho$-meson screening masses
 at $T/T_{\rm pc}=0.93$.
 The Fourier series is used for the fitting.
 Here, $m_{\rho}/T$ is multiplied by 0.85.
 }
 \label{fit_scr_093}
\end{figure}
%%%%%%%%%%%%%%%

The coefficients obtained from the $\chi^2$ fitting
are tabulated in Table~\ref{TABLE_Fourier-fit},
together with the value of $\chi^2$ degree of freedom (dof).
For both pion and $\rho$-meson,
the values of $a^{(2)}_{\textrm{F},i}$ have large error bars,
indicating that the
$a^{(2)}_{\textrm{F},i}$ cannot be determined precisely
from the present LQCD data.
Hence, we use $G^{1}_{\textrm{F},i}$ only for the extrapolation
at $T/T_{\rm pc}=0.93$.
In Fig.~\ref{fit_scr_093}, we plot the fitting result
in which two lines correspond to the upper and lower 
bounds of fitting.

Next, we consider the cases of $T/T_{\rm pc}=1.08$ and 1.35.
The polynomial series $G^{n}_{\textrm{P},i}$ is used
for the fitting.
Indeed, the $m_{i}(T,\theta)/T$ in Fig.~\ref{meson_screening_mass_mui}
are monotonically decreasing.
This suggests that
the polynomial fitting works well.
As the fitting functions,
we take $G^{1}_{\textrm{P},i}$ and $G^{2}_{\textrm{P},i}$ for $T/T_{\rm pc}=1.08$
and $G^{1}_{\textrm{P},i}$, $G^{2}_{\textrm{P},i}$ and $G^{3}_{\textrm{P},i}$
for $T/T_{\rm pc}=1.35$.

%%%%%%%%%%%%%%%%
%%    Fig. 5
%%%%%%%%%%%%%%%%
\begin{figure}[b]
 \begin{center}
  \includegraphics[width=0.4\textwidth]{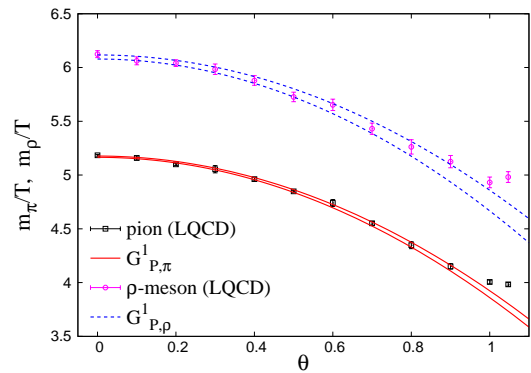}
 \end{center}
 \vspace{-10pt}
 \caption{
 The polynomial fitting results of pion and $\rho$-meson screening masses
 at $T/T_{\rm pc}=1.08$.
 Note that the fitting is performed
 by using the data only in $0\le \theta\le 0.8$.
 }
 \label{fit_scr_108}
\end{figure}
%%%%%%%%%%%%%%%

%%%%%%%%%%%%%%%%%
%%%%% TABLE 3
%%%%%%%%%%%%%%%%%
\begin{table*}[t]
 \begin{center}
  \caption{
  Coefficients of the polynomial series at $T/T_{\rm pc}=1.35$.
  }
  {\tabcolsep = 5mm
  \begin{tabular}{c|cccccc}
   \hline \hline
   meson & $T/T_{\rm pc}$ & $a^{(0)}_{\textrm{P},i}$
           & $a^{(1)}_{\textrm{P},i}$ & $a^{(2)}_{\textrm{P},i}$
                   & $a^{(3)}_{\textrm{P},i}$ & $\chi^2/{\rm dof}$ \\
   \hline
   pion & 1.35 & 5.633(6) & $-1.208$(16) & ----- & ----- & 4.332 \\
        & 1.35 & 5.660(7) & $-1.473$(49) & 0.2950(520) & ----- & 1.203 \\
        & 1.35 & 5.669(9) & $-1.641$(99) & 0.7804(2530) & $-0.3286$(1677) & 0.905 \\
   $\rho$-meson & 1.35 & 6.179(11) & $-1.316$(26) & ----- & ----- & 2.932 \\
                & 1.35 & 6.213(13) & $-1.711$(86) & 0.4119(868) & ----- & 0.695 \\
                & 1.35 & 6.210(14) & $-1.644$(191) & 0.2084(5233) & 0.1382(3506) & 0.762 \\
   \hline \hline
  \end{tabular}
  }
  \label{TABLE_polynomial-fit2}
 \end{center}
\end{table*}
%%%%%%%%%%%%%%%

For the case of $T/T_{\rm pc}=1.08$,
the coefficients and the $\chi^2/\textrm{dof}$
after the fitting are summarized
in Table~\ref{TABLE_polynomial-fit1}.
In this case,
the data are taken only in the range $0\le \theta\le 0.8$,
as already mentioned above.
The errors of $a^{(2)}_{\textrm{P},i}$ have the same order
as the corresponding mean values,
although the $\chi^2/\textrm{dof}$ is slightly improved
for $\rho$-meson.
The coefficient $a^{(2)}_{\textrm{P},i}$
cannot be determined clearly from the present LQCD data,
and hence we take $G^{1}_{\textrm{P},i}$ only as a
good fitting function for $T/T_{\rm pc}=1.08$,
and extrapolate the $G^{1}_{\textrm{P},i}$ to the $\mu_{\rm R}/T$ region.
Figure~\ref{fit_scr_108} shows the fitting result in which
the upper and lower bounds of fitting are also plotted.

%%%%%%%%%%%%%%%%
%%    Fig. 6
%%%%%%%%%%%%%%%%
\begin{figure}[t]
 \begin{center}
  \includegraphics[width=0.4\textwidth]{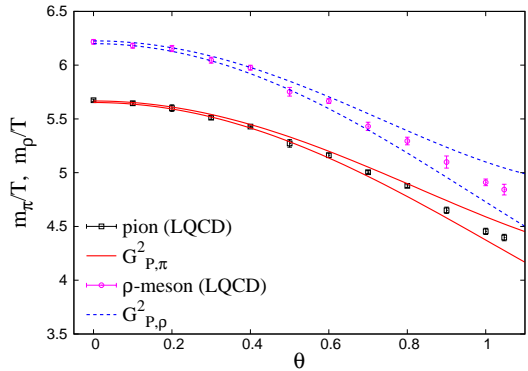}
 \end{center}
 \vspace{-10pt}
 \caption{
 The polynomial fitting results of pion and $\rho$-meson screening masses
 at $T/T_{\rm pc}=1.35$.
 }
 \label{fit_scr_135}
\end{figure}
%%%%%%%%%%%%%%%

Table~\ref{TABLE_polynomial-fit2} presents
the obtained coefficients and the $\chi^2/\textrm{dof}$
for $T/T_{\rm pc}=1.35$.
From Table~\ref{TABLE_polynomial-fit2},
the errors of $a^{(3)}_{\textrm{P},i}$
are large and same order of
the corresponding mean values.
In addition,
the value of $\chi^2/\textrm{dof}$
is considerably improved
when the $G^{2}_{\textrm{P},i}$ is used
for the fitting function,
instead of the $G^{1}_{\textrm{P},i}$.
Hence,
we use the $G^{2}_{\textrm{P},i}$
as an extrapolation function for $T/T_{\rm pc}=1.35$.
The fitting result is presented
in Fig.~\ref{fit_scr_135}.

%%%%%%%%%%%%%%%%%%%%%%%%%%%%%%%%%%%%%%%%%%%%%%%%%%
%%%%%%%%   Meson screening masses at $\mu_{\rm R}/T$
%%%%%%%%%%%%%%%%%%%%%%%%%%%%%%%%%%%%%%%%%%%%%%%%%%
\subsection{Meson screening masses at $\mu_{\rm R}/T$}
\label{extrapolation_mur}

%%%%%%%%%%%%%%%%
%%    Fig. 7
%%%%%%%%%%%%%%%%
\begin{figure}[t]
 \begin{center}
  \includegraphics[width=0.4\textwidth]{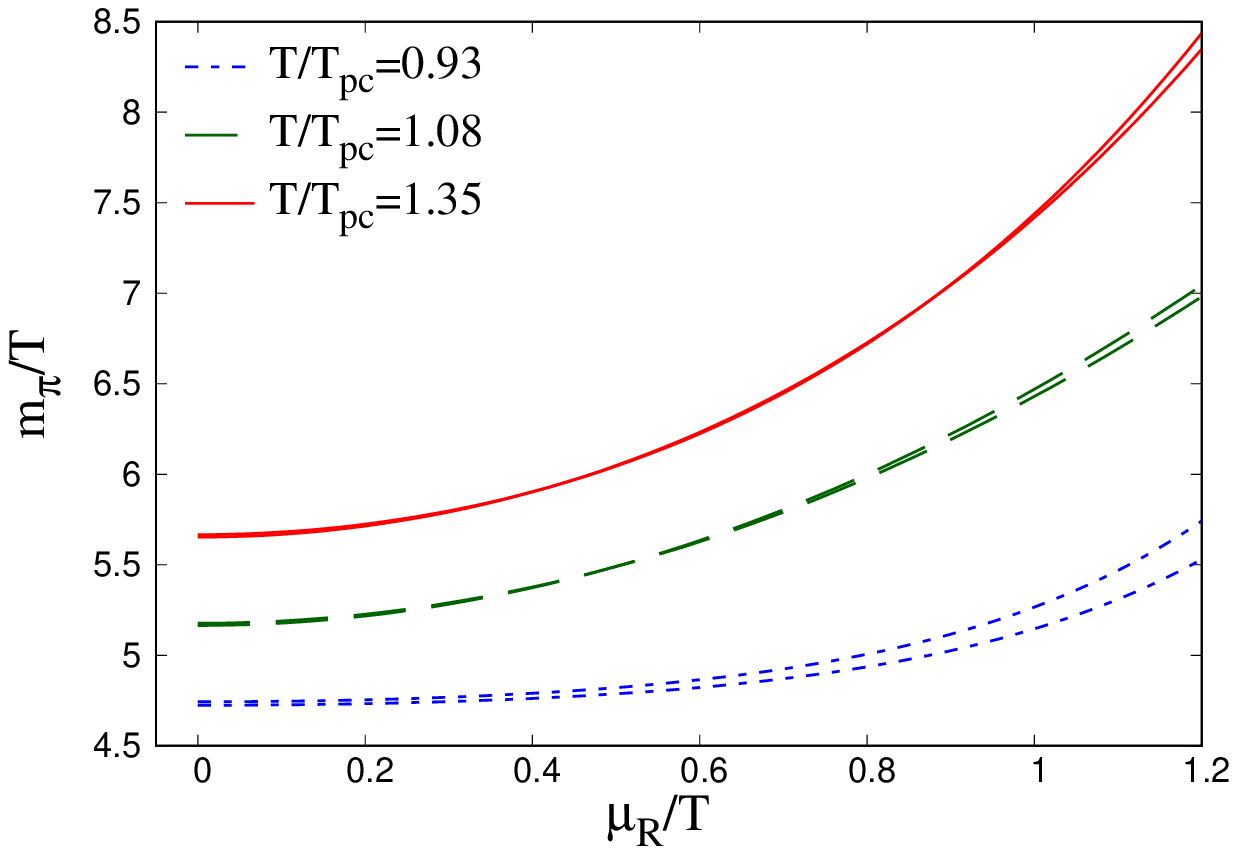}
  \includegraphics[width=0.4\textwidth]{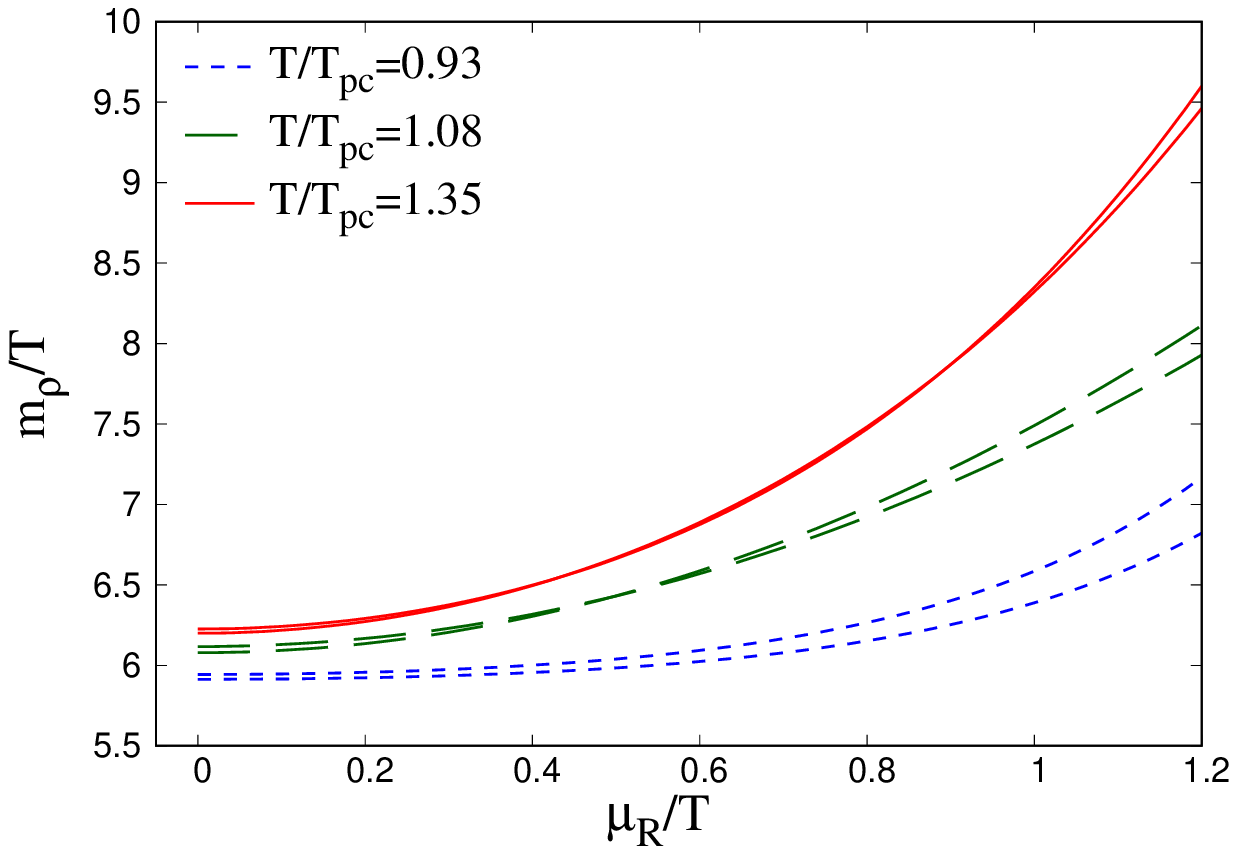}
 \end{center}
 \vspace{-10pt}
 \caption{
 The $\mu_{\rm R}/T$ dependence of the
 pion and $\rho$-meson screening masses.
 The functional form is represented in
 Eqs.~(\ref{mu_R-Fourier}) and (\ref{mu_R-Polynomial}).
 }
 \label{mur-dependence_scr}
\end{figure}
%%%%%%%%%%%%%%%

After the replacement $\theta\rightarrow -i\mu_{\rm R}/T$,
we can obtain the meson screening masses
at finite $\mu_{\rm R}/T$;
as for the extrapolated functional form,
see Eqs.~(\ref{mu_R-Fourier}) and (\ref{mu_R-Polynomial}).
Figure~\ref{mur-dependence_scr}
shows the resulting $\mu_{\rm R}/T$ dependence of
pion and $\rho$-meson screening masses
for three temperatures.
At $\mu_{\rm R}/T=0$,
pion screening mass is sensitive to temperature,
while $\rho$-meson screening mass is not.
This result is reasonable since
$\rho$-meson is heavier than pion.
In addition, as $\mu_{\rm R}/T$ increases,
pion and $\rho$-meson screening masses are monotonically increasing
in all the temperatures.

In our previous work~\cite{takahashi2} on the quark number density,
the result of the imaginary-$\mu$ approach
is consistent with the one obtained from
the Taylor expansion method
up to $\mu_{\rm R}/T \sim 0.8$ at low temperature.
In this range,
the screening masses are
almost constant for $T/T_{\rm pc}=0.93$,
and this indicates that the chiral
symmetry is not restored.
Meanwhile,
the $\mu_{\rm R}/T$ dependence
is remarkable for $T/T_{\rm pc}=1.08$ and $1.35$
because of the chiral symmetry restoration.

%%%%%%%%%%%%%%%%%%%%%%%%%%%%%%%%%%%%%%%%%%%%%%%%%%
%%%%%%%%  Summary 
%%%%%%%%%%%%%%%%%%%%%%%%%%%%%%%%%%%%%%%%%%%%%%%%%%
\section{Summary}
\label{Summary}
In this paper,
we investigated the $\mu$ dependence
of pion and $\rho$-meson screening masses 
in the imaginary and real regions 
by using LQCD simulations. 
Meson correlators were calculated at imaginary $\mu$ 
by using LQCD simulations on an $8^{2}\times 16\times 4$ lattice 
with the clover-improved two-flavor Wilson fermion action 
and the renormalization-group-improved Iwasaki gauge action.
Next, the meson correlators were fitted by
the exponential form at large $z$,
and thereby
the meson screening masses were extracted
as a function of $\theta$ for $T/T_{\rm pc}=0.93, 1.08$
and $1.35$; 
note that the system is in the confinement (deconfinement) phase 
for $T/T_{\rm pc}=0.93~(1.35)$ 
and at $T/T_{\rm pc}=1.08$ it is in 
the confinement phase for small $\theta$ and 
the deconfinement phase for large $\theta$. 

To obtain the $\mu_{\rm R}/T$ dependence
of the screening masses by the analytic continuation,
we fitted the LQCD data by the analytic function;
the Fourier series was used for the case of $T/T_{\rm pc}=0.93$,
while the polynomial series was applied
for $T/T_{\rm pc}=1.08, 1.35$. 
From the fitting,
we found that
the higher-order contributions become
important as $T$ increases.
Finally, the meson screening masses
are extrapolated to the $\mu_{\rm R}/T$ region
by the replacement $\theta\rightarrow -i\mu_{\rm R}/T$. 
It is found that, for all the $T$ we took, 
the pion and $\rho$-meson screening masses
are monotonically increasing as 
 $\mu_{\rm R}/T$ becomes large.

%%%%%%%%%%%%%%%%%%%%%%%%%%%%%%%%%%%%%%%%%%%%%%%%%%%%%%%%%%%%%%%%%%%%%%%%%%%%%%%%
%%%%% Acknowledgments 
%%%%%%%%%%%%%%%%%%%%%%%%%%%%%%%%%%%%%%%%%%%%%%%%%%%%%%%%%%%%%%%%%%%%%%%%%%%%%%%%
\noindent
\begin{acknowledgments}
 We thank A. Nakamura and K. Nagata for useful discussions
 and giving the LQCD program codes.
 J.S., H. K., and M. Y. are supported by
 Grant-in-Aid for Scientific Research
 (No. 27-7804, No. 26400279, No. 17K05446, and No. 26400278) from
 the Japan Society for the Promotion of Science (JSPS).
 The numerical calculations were performed on NEC SX-ACE at CMC, Osaka University.
\end{acknowledgments}

%%%%%%%%%%%%%%%%%%%%%%%%%%%%%%%%%%%%%%%%%%%%%%%%%%%%%%%%%%%%%%%%%%%%%%%%%%%%%%%%
%%%%% References 
%%%%%%%%%%%%%%%%%%%%%%%%%%%%%%%%%%%%%%%%%%%%%%%%%%%%%%%%%%%%%%%%%%%%%%%%%%%%%%%%

\end{document}